\documentclass[12pt]{article}
\usepackage{pdfpages}
\def\nl{\hfill\break}       
\def\newpage{\vfill\eject}       
\def\np{\vfill\eject}       
\def\ni{\noindent}

\def\cl{\centerline}
\def\part{\partial}
\font\tenbfsy=cmbsy10
\font\sevenbfsy=cmbsy10 at 7pt
\font\fivebfsy=cmbsy10 at 5pt
\newfam\bfsyfam

\textfont\bfsyfam=\tenbfsy
\scriptfont\bfsyfam=\sevenbfsy
\scriptscriptfont\bfsyfam=\fivebfsy

\skewchar\tenbfsy='60 \skewchar\sevenbfsy='60 \skewchar\fivebfsy='60
\font\tenmitb=cmmib10
\font\sevenmitb=cmmib10 scaled \magstep0 
\font\fivemitb=cmmib10 at 5pt
\newfam\mitbfam
\textfont\mitbfam=\tenmitb
\scriptfont\mitbfam=\sevenmitb
\scriptscriptfont\mitbfam=\fivemitb
\skewchar\tenmitb='177 \skewchar\sevenmitb='177 \skewchar\fivemitb='177

\font\tenhbo=cmr10 
\newfam\hbofam
\textfont\hbofam=\tenhbo
\def\hbo{\fam\hbofam}
\catcode`@=11
\def\hexnumber@#1{\ifnum#1<10 \number#1\else
 \ifnum#1=10 A\else\ifnum#1=11 B\else\ifnum#1=12 C\else
 \ifnum#1=13 D\else\ifnum#1=14 E\else\ifnum#1=15 F\fi\fi\fi\fi\fi\fi\fi}
\def\bffam@{\hexnumber@\bffam}
\def\mitbfam@{\hexnumber@\mitbfam}
\def\bfsyfam@{\hexnumber@\bfsyfam}
\mathchardef\ctimes="2\bfsyfam@02 
\mathchardef\nabl="0\bfsyfam@72   
\mathchardef\outpr="0\bfsyfam@5E  
\mathchardef\vx="0\mitbfam@18     
\mathchardef\vkappa="0\mitbfam@14     
\mathchardef\vbeta="0\mitbfam@0C  
\mathchardef\bDelt="0\mitbfam@01  
\def\fhat{\mathaccent"0\bffam@5E}  

\catcode`@=12
\mathchardef\Delt="0101           
\def\tens#1{{\hbo #1}}

\def\smalll#1{{\scriptscriptstyle{#1}}}
\def\inpr{\raise1.3pt\hbox{$\smalll\bullet$}}       
\def\iinpr{\!\mathbin{\raise2.4pt\hbox{$\smalll    
          {\smalll\bullet\atop\smalll\bullet}$}}\!}
\font\bbf=cmr12 
\font\sbbf=cmr10 

\def\section#1{\vskip5truemm\leftline{\bbf#1}\vskip 0 truemm}
\def\subsection#1{\vskip2truemm\leftline{\sbbf#1}\vskip 0 truemm}
\def\({\bigl(}
\def\){\bigr)}
\def\^#1{^{(#1)}}
\def\hoog#1{\raise1pt\hbox{$#1$}}
\def\laag#1{\raise-2pt\hbox{$#1$}}
\def\pd(#1,#2){\tf(\partial#1/\partial#2)}
\def\sf(#1/#2){{\laag{#1}\over\hoog{#2}}}
\def\tf(#1/#2){\displaystyle\sf(#1/#2)}
\def\f(#1/#2){\hbox{$#1\over#2$}}

\def\ep#1{$\epsilon-pseudo#1$}

\def\eps{\epsilon}
\def\epsbr{\overline{\epsilon}}
\def\xbf{{\bf x}\,}
\def\Abf{{\bf A}\,}

\def\DA{{\cal D}({\Abf})\,}
\def\bv{{\bf b}\,}
\def\jv{{\bf j}\,}
\def\kv{{\bf k}\,}
\def\uv{{\bf u}\,}
\def\vv{{\bf v}\,}
\def\xv{{\bf x}\,}
\def\Bbf{{\bf B}\,}

\def\Fbf{{\bf F}\,}

\def\Ibf{{\bf I}\,}
\def\Jbf{{\bf J}\,}

\def\Lbf{{\bf L}\,}
\def\Mbf{{\bf M}\,}

\def\Vbf{{\bf V}\,}

\def\ubf{{\bf u}\,}

\begin{document}
\centerline{\bf PSEUDO-SPECTRUM OF THE RESISTIVE MAGNETO-HYDRODYNAMICS}
\centerline{\bf OPERATOR: RESOLVING THE RESISTIVE ALFV\'EN PARADOX}
\vskip1true mm
\centerline{\bbf }
\vskip3true mm
{\bf
\cl{D.Borba$^*$}
\cl{$^*$JET Joint Undertaking, Abingdon, Oxon, OX14 3EA, UK}
\cl{K.S.Riedel$^{\dag}$}
\cl{$^{\dag}$ New York University, 251 Mercer St., New York  NY 10012-1185}
\cl{W.Kerner$^*$, G.T.A.Huysmans$^*$,M.Ottaviani$^*$,}
\cl{$^*$JET Joint Undertaking, Abingdon, Oxon, OX14 3EA, UK}
\cl{P.J.Schmid$^{**}$}
\cl{$^{**}$ Dept. of Applied Math., U. of Washington, Seattle WA 98195}
}

\section{\bf Abstract}
 The `Alfv\'en Paradox' is that as resistivity decreases, the discrete eigenmodes do not converge to the generalized
eigenmodes of the ideal Alfv\'en continuum.
To resolve the paradox, the $\epsilon$-pseudospectrum
of the  RMHD operator is considered. It is  proven that for any $\eps$, the $\epsilon$-
pseudospectrum  contains the Alfv\'en
continuum for sufficiently small resistivity.
Formal $\epsilon-pseudoeigenmodes$ are constructed  using the formal
Wentzel-Kramers-Brillouin-Jeffreys solutions, and
it is shown that the entire stable half-annulus
of complex frequencies with
$\rho{|\omega|^2}=|\kv \cdot \Bbf(x)|^2$ is
resonant to order $\epsilon$, i.e.~belongs to the $\epsilon-pseudospectrum$.
The resistive eigenmodes are exponentially ill-conditioned as a basis and
the condition number is proportional to $\exp(R_M^{1\over 2})$, where $R_M$ is the magnetic Reynolds number.

{\bf Keywords}:
resistive magnetohydrodynamics, pseudospectrum, non-normal operators,
continuous spectrum, Alfv\'en waves,  magnetohydrodynamic stability.

\
PACS  03.40.Kf, 47.65.+a, 52.30.Jb, 52.35.Bj

\np
\section{\bf I. INTRODUCTION}

\

In  magnetohydrodynamics (MHD), Alfv\'en
waves are represented as continuous spectra of
the linear MHD operator$^{1-5}$, where every field line oscillates with its own
frequency given by $\omega_a(x)={\bf k} \cdot {\bf B}(x)$.
Alfv\'en wave heating is based on resonant absorption by phase-mixing
at the Alfv\'en resonance$^{3,6}$.
When resistivity is included in  the linear MHD equations, the Alfv\'en
continuum is replaced by a discrete  set of eigenmodes$^{7-11}$.

One would naively expect that the normal-mode analysis of dissipative
MHD would converge
to the ideal spectrum in the limit of asymptotically small resistivity.
As the resistivity, $\eta$, decreases, the distance between eigenfrequencies
decreases as $ \eta^{1/2}$. 
 The resistive
eigenvalues lie on specific curves in the stable frequency half-plane, and
these curves are independent of resistivity for small resistivity.
{\it The resistive magntohydrodynamics (RMHD) paradox is that
the resistive eigenmodes
do not converge to the ideal continuum as the resistivity becomes
vanishingly small.}

To resolve this paradox,
we consider the $\epsilon-pseudospectrum^{12-17}$, 
a generalization of
the spectrum which corresponds to approximate eigenmodes.
We show that for any $\eps$, the $\epsilon$-pseudospectrum of resistive MHD
contains the continuous spectrum of ideal MHD
for sufficiently small values of the resistivity, $\eta$.
Using the Wentzel-Kramers-Brillouin-Jeffreys (WKBJ) approximation$^{8-11}$,
we show that the entire
half-annulus, $\rho |\omega|^2 = |\kv\cdot \Bbf(x)|^2$, Im$[\omega] >0$,
is contained in the $\epsilon$-pseudospectrum with the critical value of
$\eps$ required for the existence of a $\epsilon$-eigenmode, $\eps_{crit} \sim \exp(-1/\eta^{1\over 2})$.

Since the resistive spectrum and the ideal spectrum
are different, we examine the question: ``Which spectrum is more relevant
in describing the time evolution on the ideal MHD time scale?''
Perturbations in ideal MHD decay algebraically due to phase mixing.
If the resistive MHD eigenmodes form a complete basis, then one would
expect that initial
perturbations would decay exponentially. The strong damping of the resistive
eigenmodes has caused authors$^{9,10}$ to question the completeness
of the resistive spectrum and the significance
of the resistive spectrum. For a similar problem in fluid dynamics,
it has been shown that the Orr-Sommerfeld
equations have a complete set of eigenmodes$^{18}$.
 Therefore, it is reasonable to believe that the resistive MHD eigenmodes
are also complete. We show that the resistive eigenmodes are strongly
non-orthogonal and that the condition number of the RMHD eigenfunction basis
degrades exponentially with the square root of the magnetic Reynolds number.
Consequently, expanding an arbitrary initial perturbation in eigenmodes
gives an ill-conditioned representation of the time evolution
until times of order $O(R_M^{1\over 2})$,
where $R_M$ is the magnetic Reynolds number.


In Section II, the resistive MHD equations are presented.
In Section III, the $\epsilon$-pseudospectrum is defined. In Section IV, we
show that
for small resistivity, the continuous spectrum of ideal MHD is contained in
$\epsilon$-pseudospectrum of resistive MHD.
In Section V, we analyze the $\epsilon$-pseudospectrum using the WKBJ
expansion. Section VI presents our numerical results.
Section VII considers representations of the initial value problem in terms
of the RMHD eigenmodes. Section VIII discusses the
transient growth problem$^{12-15,17,19}$.
Section IX summarizes our findings.
Appendix A gives the appropriate generalizations of $\eps$-pseudospectra
to the generalized eigenvalue problem. Appendix B evaluates the WKBJ
phase integral for the linear profile. Appendix C presents our finite-element
discretization. In Appendix D, the WKBJ approximation for the
$\eps$-pseudospectrum is presented. Appendix E shows that transient growth
occurs in ideal MHD when the initial perturbation is tearing mode-like.

\section{\bf II. LINEAR MAGNETOHYDRODYNAMICS}

\

We denote the equilibrium magnetic field by $\Bbf(\xv)$ and
the equilibrium current  by $\Jbf(\xv)$.
We consider incompressible MHD equations linearized about a no-flow
equilibrium ($\Vbf{(\xv)} \equiv 0$) with constant density, $\rho\equiv 1$:
$$
\rho {\pd({\bf v},t)}={\bf J} \times {\bf b} +( \nabla \times {\bf b}) \times
{\bf B} - \nabla p
\eqno(1a)$$
   $${\pd({\bf b},t)}=\nabla \times ({\bf v} \times {\bf B} - \eta \nabla
\times {\bf b})
\eqno(1b)$$
$$ \nabla \cdot {\bf v}=0
\eqno(1c)$$
$$ \nabla \cdot {\bf b}=0 \ ,\eqno(1d)$$
where $\bf v$, ${\bf b}$, and $p$ denote the flow velocity, magnetic field
and pressure perturbations. We consider time dependent perturbations
of the form
$\bv({\bf x},t)=e^{\lambda t}\bv({\xv})$, and define $\omega = -i \lambda$.
We write the resulting eigenvalue problem symbolically as
$\Lbf_{\eta} U = \lambda U$, where U is the state vector:
$U \equiv (\vv,\bv)^T$ and $\Lbf_{\eta}$ is the resistive MHD operator.

In the  absence of resistivity, the $\delta W$ energy principle
shows that ideal MHD is a self-adjoint operator$^{1,2,5}$. (More precisely,
when the perturbed magnetic field is eliminated, Eq.~(1) is rewritten
as $\Fbf \xi = \rho \omega^2 \xi$,where $\xi \equiv {{\bf v} / \lambda}$
and $\Fbf$ is a symmetric operator$^1$ in $\omega^2$.
In Ref.~5, it is shown that $(\Fbf + \sigma \Ibf)$ is self-adjoint
where $\sigma \Ibf$ is a multiple of the identity operator which
makes  the combined  operator positive.) Note that the ideal MHD operator is
self-adjoint only when the initial perturbation has the form:
$\vv({\bf x},t=0)= {\bf \xi},\ $
$\bv({\bf x},t=0)= \nabla \times ({\bf \xi} \times {\bf B})$

Since the ideal MHD operator is normal,
an initial perturbation
can be represented
as a sum over the discrete eigenmodes plus an integral over the
generalized eigenfunctions of the continuous spectrum.
The eigenfunctions have a ${1\over x}$ singularity at the resonant point where
$\omega= \pm \kv \cdot {\Bbf}(x) $.
In contrast, linear resistive MHD is not a normal system of equations,
and thus, the eigenmodes need not be orthogonal or even form a complete
basis. In Sections IV and V, we show that the resistive eigenmodes
are strongly non-orthogonal.


\section{\bf III.  EPSILON-PSEUDOSPECTRUM}

\

The spectrum of a linear operator corresponds to complex frequencies
where the frequency response Green's function is infinite, and these
frequencies dominate the long time asymptotics.
The finite-time evolution of a non-normal operator can be significantly
modified by frequencies where the Green's function
is large, but not infinite. Therefore, we consider a generalization of the
spectrum to near-resonance, the $\epsilon$-pseudospectrum:

\noindent
Definition 1 [Ref.~12].{\it
Let $\Abf$ be a  closed linear operator with domain, $\DA$,
and let $\epsilon \geq 0$ be given.
A complex number $\lambda$ is in the
$\epsilon$-pseudo\-spec\-trum of ${\bf A}$,
which we denote by $\Lambda_{\epsilon}( {\bf A})$,
if one of the following equivalent conditions is satisfied:

(i) the smallest singular value of ${\bf A}-\lambda{\bf I}$ is
less than or equal to $\epsilon$.

(ii) there exists ${\bf u}\in \DA$ such that
$||{\bf u}||^2 = 1$ and
$ ||({\bf A}-\lambda{\bf I}) {\bf u}||^2 \leq \epsilon^2$,

(iii) $\lambda \in \Lambda ({\bf A})$ or
$\lambda \in \rho ({\bf A})$ and there exists ${\bf u} \in \DA$
such that $||{\bf u}||^2 = 1$
and ${\bf u}^{*} ({\bf A}-\lambda{\bf I})^{-1 *}$
$({\bf A}-\lambda{\bf I})^{-1} {\bf u} \geq 1/
\epsilon^2$,

(iv) $\lambda$ is in the spectrum of
${\bf A}+ \epsilon  {\bf E}$,
where the operator ${\bf E}$ satisfies
$\parallel {\bf E} \parallel \leq 1$.
}

The stated definition is for finite dimensional operators. For
infinite dimension problems, we need to extend these definitions
by replacing $\uv$ with a sequence of functions, $\{ \uv_n \}$, in the
domain of  $\Abf$, i.e.~require that Definition 1 hold on
the closure of the  domain of $\Abf$. Thus, condition (iii) becomes
$||({\bf A}-\lambda{\bf I})^{-1} || \geq 1/\epsilon$.

We measure the size of the operator and of the perturbation in the
operator norm; i.e.~ 
$||{\bf A}||\equiv sup_{\bf u}{||{\bf A u}||\over  {||\bf u}||}$
. Since the operator norm depends on the norm of
the underlying function space, so does the definition
of the $\epsilon$-pseudospectrum.
For resistive MHD, we use the energy norm,
$\int \left( |\vv|^2 +  |\bv| ^2 \right) dV $. 
In a coordinate system where $||\uv||^2$ is equal to the $L_2$ inner product,
the operator norm
corresponds to the largest singular value of the matrix representation
of $\Abf$.
We denote the smallest singular value of ${\bf A}-\lambda{\bf I}$ by
$\eps_{bd}(\lambda)$.

We call the test function in part (ii),  $\uv$,  an $\eps$-pseudomode
and call $\lambda$  an $\epsilon-pseudoresonance$. Part (iii) states that
$\lambda \in \Lambda_{\epsilon}( {\bf A})$ is equivalent to the norm of
the frequency response Green's function at $\lambda$
being size ${1\over \eps}$ or larger.
Part (iv) says that the operator can be perturbed by a term of size
$\eps$ such that the modified operator has an exact resonance at $\lambda$.

The equivalence of these four conditions is proven in Ref.~12, 13 and 16.
In several excellent articles, the $\epsilon$-pseudospectrum is analyzed for
the Orr-Sommerfeld  equation$^{12-15}$  and the convection diffusion
equation$^{17}$.

In resistive MHD, the simple eigenvalue problem, $\Abf \uv =\lambda \uv$, is
replaced by the generalized eigenvalue problem, $\Abf \uv =\lambda \Mbf \uv$,
where the weight matrix, $\Mbf$, is the matrix defined in the energy norm
of the perturbation. In Appendix A, we give generalizations
of Definition 1 to the generalized eigenvalue problem.
When $\bf M$ is self-adjoint, positive definite and bounded
above and below, we can transform the
problem into a standard eigenvalue problem: ${\bf A}_{\bf F}
{\bf u}^{\prime} =  \lambda{\bf u}^{\prime}$
where ${\bf F}^{*} {\bf F} = {\bf M}$,  ${\bf u}' = {\bf F}{\bf u}$, and
${\bf A_F} \equiv{\bf F}^{*-1 } {\bf A}{\bf F}^{-1}$,
where $^*$ denotes the adjoint operator.
We then compute the $\epsilon$-pseudospectrum of the standard linear problem.
This transformation gives the $\epsilon$-pseudospectrum for the
generalized eigenvalue problem in the physically correct energy norm;
however, ${\bf F}^{*-1 } {\bf A}{\bf F}^{-1}$ is no longer a banded matrix.
As a result, the computation of the singular value decomposition is very
costly. Therefore, we consider a different generalization of
the $\epsilon$-pseudospectrum, which replaces part (i) of Definition 1 with
the smallest singular value of $\Abf - \lambda \Mbf$.
To correctly normalize this definition of the generalized
$\epsilon$-pseudospectrum, we divide the singular values of
$\Abf - \lambda \Mbf$ by $||\Mbf||$. (See Appendix A.)
We compute the boundary of the generalized $\epsilon$-pseudospectrum,
$\eps_{bd}(\lambda,\Abf,\Mbf) \equiv {1 \over ||\Mbf||} \times$
the smallest singular value of ${\Abf - \lambda \Mbf }$, as
a function of $\lambda$.

\

\section{\bf IV. EPSILON-PSEUDOMODES AND SINGULAR SEQUENCES }

\

As a first step in resolving the Alfv\'en paradox,
we show that for sufficiently
small $\eta$, the continuous spectrum of ideal MHD is contained in the
$\epsilon$ pseudospectrum. 
We begin by stating a lemma$^{20-21}$ on singular sequences:

\noindent
Lemma 1: {\it The spectrum of a self-adjoint operator, $\Lbf$,
consists of those
complex numbers, $\lambda$, for which there exists a sequence
of functions $U_n$ such that $||U_n|| =1$ and
$||\Lbf U_n  -\lambda U_n|| \rightarrow 0$. }

In Ref.~5, Laurence shows that the ideal MHD operator plus a
multiple of the identity has a self-adjoint extension.
(The multiple of the identity, $\sigma \Ibf$, is necessary to ensure
positivity of the operator.) The ideal MHD operator is self-adjoint when
the domain of the operator is restricted to
 perturbation of the form: $\vv({\bf x},t=0)= {\bf \xi},\ $
$\bv({\bf x},t=0)= \nabla \times ({\bf \xi} \times {\bf B})$, and
the mprm is $\int |\vv|^2 +  |\bv| ^2  dV $.

Since the ideal MHD operator has a self-adjoint formulation, every
value of $\lambda$ in the spectrum of the ideal MHD operator has a singular
sequence of functions. In fluid dynamics, the ideal operator is not
self-adjoint, and therefore singular sequences of functions need not exist
for the spectrum of the inviscid Orr-Somerfeld equation.
In Ref.~4, Hameiri uses singular sequences to show that ballooning modes are
part of the ideal MHD essential spectrum.

The domain of the ideal MHD operator differs from that of
the resistive MHD operator, $\Lbf_{\eta}$, because the resistive operator
involves more derivatives and requires more boundary  conditions
while the ideal MHD operator imposes the constraint that
$\bv({\bf x},t=0)= \nabla \times [\vv({\bf x},t=0) \times {\bf B}]$.
We denote the ideal MHD operator, restricted to the intersection of
the domains of the ideal and resistive MHD operators by $\Lbf_I$.
This restriction amounts to considering the ideal MHD operator on the
function space where
$|| \nabla \times \nabla \times {\bf b}||^2$ is finite.
The ideal MHD restriction:
$\bv({\bf x},t=0)= \nabla \times [\vv({\bf x},t=0) \times {\bf B}]$
remains in effect. For resistive MHD, we use the norm ,  
$\int \left( |\vv|^2 +  |\bv| ^2 \right) dV $,
while ideal MHD is self-adjoint  in a  different norm,  $\int |\vv|^2 dV $,

\ni
Definition: {\it  $\lambda$ is in the dissipative spectrum of
 the ideal MHD operator if and only if there is a singular sequence
of functions in the intersection of the
domains of the RMHD and ideal operators such that
$||U_n|| =1$ and~$||\Lbf U_n  -\lambda U_n|| \rightarrow 0$. }

By Lemma 1, the dissipative  spectrum of the ideal MHD operator
is a subset of the  spectrum of the ideal MHD operator.
In ideal MHD, the singular function sequences are usually smooth functions
which are localized near the resonance surface. In fact, we are unaware of
any spectrum in ideal MHD where the  singular sequence of test functions
cannot be constructed in the function space of the  resistive MHD operator.
We introduce the terminology of ``dissipative spectrum''
in order to prove the following theorem:

\noindent
Theorem 1: {\it Let $\lambda$ be in the dissipative  spectrum of
the ideal MHD operator, $\Lbf_{I}$,
for a bounded toroidal MHD equilibrium, and let $\epsilon >0$ be given.
Then there exists a critical value of resistivity,
$\eta_{cr}$, such that $\lambda$
is contained in the $\epsilon-pseudospectrum$ of resistive MHD
for all $0<\eta \le \eta_{cr}$.}

Proof: By Lemma 1, there is
a sequence of test functions, ${\bf U}_n$, with $||{\bf U}_n||=1$, such that
$ ||\Lbf_I{\bf U}_n -\lambda {\bf U}_n||\rightarrow 0$.
For simplicity, we denote the resistive MHD operator by
$\Lbf_{\eta}{\bf U}_n =\Lbf_{I}{\bf U}_n +
{\eta \nabla \times \nabla \times \bf {b}}_n$ where
$\bv_n$ is the magnetic field component of ${\bf U_n}$.
We apply criterium (ii) from the definition of $\eps$-spectrum.
$$
||\Lbf_{\eta}{\bf U}_n -\lambda {\bf U}_n||=||\Lbf_{I}{\bf U}_n+
{\eta \nabla \times \nabla \times \bf {b}}_n-\lambda {\bf U}_n|| \ . 
$$
Using the Minkowski inequality, it follows
$$
{||\Lbf_{\eta}{\bf U}_n -\lambda {\bf U}_n||} \le {||\Lbf_I{\bf U}_n -\lambda
{\bf U}_n||}
+\eta {||{{\nabla \times \nabla \times {\bf b}}_n}||} \ . 
$$
We select $U_n$  such that
$ ||\Lbf_I{\bf U}_n -\lambda {\bf U}_n||<{{\epsilon} \over 2} $
and select $\eta_{cr}$ such that
\nl
$\eta_{cr}  {|| {\nabla \times \nabla \times {\bf b}}}_n|| <{{\epsilon} \over 2}$. \ \ \ \ \ \  $|^-_-|$

The spectrum
of an arbitrary linear operator can be divided into three parts:
the point spectrum, the continuous spectrum and the residual spectrum.
Singular function sequences exist for the point spectrum and
the continuous spectrum, but need not exist for the residual spectrum.
Self-adjoint operators have no residual spectrum. To generalize
Theorem 1 to the inviscid Orr-Sommerfeld equation, we need to require that
$\Lbf -\lambda\Ibf$  have a singular sequence in the domain of
the viscid Orr-Somerfeld equation.

Theorem 1 is a very general result, but it is a weak result in the sense that
$\eta$ scales as $\eps$. In the next section, we derive a much stronger
scaling, $\eps \sim \exp(-1/\eta^{1\over 2})$,
for specific one-dimensional geometries.


\section
{\bf V.  WKBJ ANALYSIS}

\

We restrict ourselves to the slab geometry with coordinates,
$\xbf \equiv (x,y,z)^T$, and an equilibrium magnetic field,
$\Bbf(\xv) = \Bbf_y(x) \hat{y} +\Bbf_z(x) \hat{z}$.
We consider perturbations  of the form:
$$\bv({\bf x},t)=e^{\lambda t} e^{i (k z + m y )} \bv({x})
\ .$$
The equations can be decomposed into two separate eigenvalue
problems, the transverse eigenmode equations and the longitudinal
eigenmode problem.
Our analysis will focus on the transverse equations which describe
Alfv\'en waves. Eliminating the total pressure term in Eq.~(1a)
yields the transverse eigenmode equations$^{7-11}$:
$$
 {H(x)} \nabla^2\psi -
{H''(x)}\psi 
= \lambda \nabla^2\xi
\ , \eqno(2a)$$
$$
\eta \nabla^2\psi  + H(x) \xi =  \lambda \psi
\ , \eqno(2b)$$
where $\psi\equiv {\bf b}_x$, 
$\xi\equiv i  {\bf v}_x$, 
and $H(x) \equiv (\kv \cdot \Bbf(x))/\rho^{1\over 2}$.
We impose perfectly conducting boundary conditions at $x= x_a$ and $x=x_b$:
$ \xi(x_a)=\psi(x_a) =\xi(x_b)=\psi(x_b)=0 $.
The remainder of the article will concentrate on Eq.~(2) for a single
Fourier mode, $e^{i (k z + m y )}$.
The $\eps$-pseudospectrum of the MHD operator is the union of the
$\eps$-pseudospectra of all of the Fourier modes.

Equation (2) is a fourth order system of equations with two
formal solutions that are asymptotic to the ideal MHD solutions.
The remaining two formal solutions are constructed with
the WKBJ expansion$^{8-11}$:
$\xi \sim {(H(x)^2-\lambda^2)}^{-{{1} \over {4}}}
e^{\pm i \phi (x)}$ where  $$\phi'(x)=\sqrt{{{H(x)}^2-
{\lambda}^2} \over {i \lambda \eta}}\ .$$
The WKBJ phase function depends on the complex frequency,
$\omega \equiv -i \lambda$, and we will sometimes write  $\phi(x;\omega)$
to highlight this dependence.
The WKBJ solutions oscillate and grow exponentially with a scale-length of
$\eta^{1\over2}$.

Let $Im[\phi(x)]$ have its minimum in the
interior of the domain, at $x = x_{mn}$ with $x_a<x_{mn}<x_b$, and assume
that $Im[\phi(x_b)] \le Im[\phi(x_a)]$. (Otherwise replace $x_b$ with $x_a$
in this paragraph.) We construct a formal $\eps$-pseudomode
using the WKBJ formal solution, $e^{ i \phi (x)}$.
To satisfy the boundary conditions, we add a low order polynomial (linear term) to
$e^{ i \phi (x)}$. The size of this polynomial correction is
$O(e^{i[\phi(x_{mn})-\phi(x_{b})]})$. This formal $\eps$-pseudomode
satisfies the RMHD operator up to a perturbation of size
$\epsilon \approx O(e^{i[\phi(x_{mn})-\phi(x_{b})]})$.
Since Im$[\phi(x_{mn})-\phi(x_{b})]$ is $O({1 \over {\sqrt{\eta} }})$,
the critical value of  $\epsilon$
scales as $e^{-{1 \over {\sqrt{\eta} }}}$.

A similar construction is possible when $Im[\phi(x)]$ has
its maximum in the interior of the domain, using the WKBJ formal solution
$e^{ -i \phi (x)}$. Summarizing our results, we have

\noindent
Theorem 2: {\it
Consider a
a slab (or cylindrical) MHD equilibrium. For
sufficiently small resistivity, there is a formal WKBJ $\epsilon-
pseudoeigenmode$ of the incompressible RMHD operator
with $\eps_{bd}(\lambda) \sim \exp(-1/\eta^{1\over 2})$,
provided that $Im[\phi(x)]$ has
a strict minimum (maximum) in the interior of the domain,
i.e.~$e^{ i \phi (x)}$ as a maximum (minimum).
For monotone  ${{{\kv} \cdot {\Bbf}(x)}} \ne 0$
, the formal $\epsilon-pseudoeigenmode$ exists in the half $\lambda$-annulus,
$\rho{|\lambda|}^2=H(x)^2$, Re$[\lambda]<0$.}

To prove the last sentence of Theorem 2, we use Property 1 of Ref.~9:

\noindent
Property 1. {\it For ${\rm Re} [ \omega ] \neq 0$, ${\rm Im} [ \phi^{\prime}
(x; \omega )] = 0$ if and only if $\rho | \omega |^2 = H(x)^2$ and
Im$[\omega] \geq 0$.}

For monotone $H(x)$ with $H(x) \neq 0$, Property 1 implies that
${\rm Im } [ \phi] (x; \omega )$ has either a maximum or
a minimum in the interior of $[x_a ,x_b ]$ when $\omega$ is in the half
annulus specified in Property 1.

\

In Theorem 2, we have discussed only formal solutions.
A formal solution of Eq.~(2) can fail to be asymptotic to an actual
solution of Eq.~(2) globally. Reference 9 provides a comprehensive
discussion of the global validity of the WKBJ expansion. The formal
solutions fail because the actual solution can pick up an exponentially
growing solution, while the formal solution continues to decrease.

To construct the $\epsilon$-pseudomode in Theorem 2, we need to show that
there is an actual solution which is exponentially larger in the interior
than at the boundary.
We restrict ourselves to complex analytic $H(x)$ profiles with
a single transition
point (where $\phi^{\prime} (x) = 0$) in the complex domain around the
real interval, $[x_a ,x_b ]$.
From Property 1, this restriction corresponds to
monotonically increasing  ${\kv \cdot\Bbf(x)}$ profiles with
${\kv \cdot \Bbf(x)} \ne 0$ in the domain.

The validity of the formal WKBJ solutions depends on the geometry of
level lines of  ${\rm Im } [ \phi (x)]$. 
The anti-Stokes lines are the three curves of constant ${\rm Im } [ \phi (x)]$
which emerge from the transition point.
Figure 1 displays the geometry of the anti-Stokes lines for
different regions in the complex $\lambda$-plane.

From Ref.~9, we know that when one or no anti-Stokes line crosses
$[x_a ,x_b ]$, the WKBJ expansion is valid. We consider the case where
two different anti-Stokes lines cross $[x_a ,x_b ]$ at $x_1 $ and $x_2$.
Without loss of generality, we assume that ${\rm Im } [ \phi (x)]$ has its
minimum at $x_{mn}$ with $x_a < x_1 < x_{mn} < x_2 < x_b$. From Ref.~9,
the WKBJ expansion $e^{i \phi (x)}$ is valid in $[x_a ,x_2 ]$,
but  we cannot exclude the possibility that the actual solution is not
asymptotic to $e^{i \phi (x)} + ce^{2i \phi (x_2 ) -i \phi (x)}$ in the
interval $[x_2 ,x_b ]$, i.e.~the actual solution grows exponentially
in $[x_2 ,x_b ]$. To construct an $\epsilon$-pseudomode, we need to require
that $| e^{i \phi (x_{mn})} | \gg |e^{2i \phi (x_2 )-i \phi (x_b )} |$
or in other words ${\rm Im} [ \phi (x_{mn} ) + \phi (x_b ) -2 \phi (x_2 )]
< 0$. Alternatively, we can use the WKBJ solution in $[x_1 ,x_b ]$ and
analytically continue it in $[x_a ,x_1 ]$. In this case, the
$\epsilon$-pseudomode construction is successful if ${\rm Im} [ \phi (x_{mn} )
+ \phi (x_a ) - 2 \phi (x_1 )] < 0$.

In summary, our results are:

\noindent
Theorem 3. {\it Let $H (x)$ be a complex analytic, monotonic profile with
$H(x) \neq 0$ in $[x_a ,x_b ]$ and a single transition point in the complex
region around $[x_a ,x_b ]$. There is an actual $\epsilon$-pseudomode of the incompressible RMHD operator
with $\epsilon \ \sim \ \exp (- \eta^{1/2} )$ for complex eigenfrequencies
which satisfy Im$[\omega] < 0 $, $\rho | \omega |^2 = H(x)$, for a value
in $[x_a ,x_b ]$, and one of the two conditions:
$$
{\rm Im} [ \phi (x_{mn} ) + \phi (x_b ) -2 \phi (x_2 )]
< 0
\ , \eqno(3a)$$ $$
{\rm Im} [ \phi (x_{mn} ) + \phi (x_a ) - 2 \phi (x_1 )] < 0
\ , \eqno(3b)$$
where ${\rm Im} [ \phi]$ has a minimum at $x_{mn}$, and the anti-Stokes lines
are located at $x_1$ and $x_2$.}

Our numerical results indicate that
$\epsilon$-pseudomodes exist with $\epsilon \leq \exp ({-{{1} \over {\eta^{1/2}}}})$
even when the conditions of Eq.~(3) are not fulfilled.
This frequency region is near the ideal MHD continuous spectrum.
Thus it may be possible to construct $\epsilon$-pseudomodes
in this region using a combination of the WKBJ formal solutions
and the ideal MHD formal solutions.

\section{\bf VI. NUMERICAL RESULTS}

Figure 2 shows the $\epsilon -pseudospectrum$ contours for Alfv\'en
waves,
computed by applying the singular value decomposition to $\Abf -\lambda \Mbf$.
A linear slab equilibrium with $H(x) = x$, $x_a = 0.2$, $x_b = 0.4$,
and $\rho =1$ is used.
The Alfv\'en frequency, $\omega_A$, varies linearly from 0.2 to 0.4.
The finite-element discretization is described in Appendix C.
The number of radial cubic finite-elements used in the discretization
of the equation is 41, which is sufficient to resolve
the structure of the solutions for $\eta=10^{-4}$.
In the same figure, we superimpose  the resistive MHD eigenvalues.

Table 1 displays the RMHD eigenvalues for the  finite-element discretization
and the WKBJ approximation. For the WKBJ approximation,
we have used the dispersion relation: $\phi(x_a)-\phi(x_b)= n \pi $.
The good agreement between the numerical and the WKBJ  eigenvalues
demonstrates the accuracy of our numerical method.

Figure 3 compares the numerical and analytical contours of the
$\epsilon$-pseu\-do\-spec\-trum.
The solid lines are the computed values of $\eps_{bd}$ and
the dashed lines correspond to the WKBJ approximation presented in Appendix D (D9).
The large $\epsilon$-pseudospectrum contours around the triple point
indicate that these eigenvalues are very  sensitive to perturbations.
The similarity of the  numerical and WKBJ contours show that
the  eigenvalue sensitivity is
not due to the numerical discretization of the original equations,
but rather is a property of the RMHD  operator.

The discrepancy between the analytic and numeric $\epsilon$-pseudospectra
occurs in the $\lambda$ region which is to the right of the triple point and which
has two  anti-Stokes lines intersect $[x_a,x_b]$. The calculation
of the  analytic $\epsilon$-pseudospectrum in Appendix D explicitly
assumes that the WKBJ solutions are valid globally, and this assumption
fails when two  anti-Stokes lines intersect $[x_a,x_b]$.
Thus, it is natural that  some discrepancy occurs below the
triple point in Fig.~3.

In Figure 4, we present a cross-section of the $\epsilon-pseudospectrum$
at $Re[\lambda]=-0.1$ for different values of the resistivity.
 A comparison between the analytical and the numerical
$\epsilon-pseudospectrum$ is done in Figure 5, for $\eta=10^{-4}$.

The most sensitive eigenvalues are those which are located near
the triple point where the three branches of the different eigenvalues curves come together. The
\ep{spectrum} contours expand rapidly around the triple point.

If the finite numerical accuracy is smaller than  $\epsilon$ and the
corresponding $\epsilon-pseudospectrum$
contour extends into a large region,
then the numerical code cannot properly resolve the eigenvalues
inside the $\epsilon-pseudospectrum$ contour,
regardless of the number of grid points which are used in the discretization.
For very small resistivity, the $\epsilon-pseudospectrum$ is lower than
machine roundoff  in the half-annulus given in Theorem 2, 
and the resulting eigenvalues are
scattered within this region. 

\section{\bf VII. INITIAL VALUE PROBLEM USING RMHD EIGENMODES}

For the Orr-Sommerfeld equation,
a similar problem in fluid dynamics, it has been shown that the
eigenmodes form a complete basis$^{18}$.
Therefore, it is reasonable to believe that the resistive MHD eigenmodes
are complete as well.
Assuming that the resistive MHD eigenmodes form a complete basis, then
an arbitrary initial
perturbations will decay exponentially as $t \rightarrow \infty$. In contrast,
perturbations in ideal MHD decay algebraically due to phase mixing.

The strong damping of the resistive
eigenmodes has caused
authors$^{9,10}$ to question the completeness of the resistive spectrum and the
significance of the resistive spectrum.
Implicit in this argumentation is the belief that
strong exponential damping of initial perturbations
would occur on the Alfv\'enic time-scale. This intuition is based on
normal operators and expansions of the solution in orthonormal eigenfunctions.
We now show that the eigenfunctions of resistive MHD are nearly
degenerate and that the condition number of the basis is very large,
$\sim \exp( {R_M^{1/2}})$, where $R_M$ is the magnetic Reynolds number.
The extended Bauer-Fike theorem gives a lower bound on the condition number
of the eigenfunction basis in terms of the norm of the resolvent
and the distance between the complex frequency, $\omega$, and the
nearest eigenvalue. (See Appendix B of Ref.~12.) In Section IV,
we have shown that the norm of the resolvent is $O(\exp( {R_M^{1/2}}))$
in the frequency half annulus. Most of this half annulus is
a distance $O(1)$ from the eigenvalues. Thus the
Bauer-Fike lower bound on the condition number is $O(\exp( {R_M^{1/2}}))$.

In expanding an initial perturbation in the RMHD eigenfunctions,
the {\it coefficients} of the perturbation in the RMHD basis may be large,
$O(\exp( {R_M^{1/2}}))$,
due to the ill-conditioned RMHD basis.
In the eigenfunction basis, each
term individually damps on the Alfv\'en time-scale, but the coefficients are
so large and the basis is so ill-conditioned that the combined sum
behaves like the ideal solution does.
Both analytical$^{6}$ and numerical studies$^{22}$ of RMHD
have shown good agreement with ideal MHD on time scales which are long
compared to the Alfv\'en time and short relative to the resistive time.
Thus, we suggest that the resistive MHD eigenvectors are complete,
but so poorly conditioned that they should not be used to interpret the
temporal evolution on the ideal time-scale. From the bound on the
condition number of the RMHD eigenmode basis, we believe that the
eigenmode decomposition will only be useful for times of
$O(R_M^{1/2}).$

\section{\bf VIII. TRANSIENT GROWTH OF INITIAL PERTURBATIONS}

The other aspect of temporal evolution generated by non-normal operators
is {\it transient growth}. Transient growth occurs when an initial perturbation
grows in magnitude, as measured by the energy norm, before decaying.
When the eigenfunctions are orthogonal and complete, and the system is
stable, transient growth cannot occur. For non-normal operators, transient
growth can occur due to the non-orthogonal nature of the eigenfunction
basis. Butler and Farrell$^{19}$ and Reddy, Schmid and Henningson$^{12,14}$
have studied transient growth in
the Orr-Sommerfeld equation and have found that initial perturbations
can be amplified by factors of thousands. The transient growth of initial
perturbations has been proposed as a mechanism by which fluctuations
reach magnitudes which trigger nonlinear instabilities.

We show below that for a constant equilibrium current  ($H(x) =x$)
the energy of the perturbation is constant, so there is no transient growth in
energy.
In Refs.~12-16 and 19, optimization algorithms are given to
determine the initial
perturbation which experiences the largest transient growth. We have applied
these algorithms to Eq.~(2) with a variety of  $H(x)$ profiles.
For these profiles, we found only limited  transient growth.

In ideal MHD, when the $\delta W$ energy principle is negative, there is
an exponentially growing instability. When the $\delta W$ is positive,
the total  energy is constant:
$$
{d \over dt} {1 \over 2} \int (|\partial_t \xi|^2 + \xi^{\dag}\Fbf\xi ) dV
=0$$
where $\Fbf$ is the $\delta W$ operator and $\xi$ is the displacement:
$\partial_t \xi ={\bf v}$.  The value of $\delta W$ is equal to the maximum
possible amplification of the kinetic energy.

This bound on the kinetic energy growth is 
valid only for perturbations of the form
$\bv(x,t =0 ) = \nabla \times (\vv(x,t =0 ) \times B)$. This restriction
corresponds to considering perturbations which only displace the flux surface
and do not change the topology of the magnetic field. In  Appendix E,
we reproduce a result of H. Grad's which shows that   linear in time growth  occurs at the rational surface
for more general perturbations with$\oint \bv \cdot \nabla p^0 d \ell$
does not vanish on a rational flux surface. Thus we expect transient growth
to be relevant for resonant perturbations with tearing mode parity.
The Grad analysis addresses only growth  in the "supremum" norm and
not with respect to the energy norm. 

We now examine the energetics of transient growth:
$$
{d\over dt} {1 \over 2} \int(|\vv|^2 + |\bv|^2 ) dV =
\int \vv \cdot (\Jbf \times \bv + \jv \times \Bbf -  \nabla p^1)
+ \bv \cdot \nabla \times (\vv \times \Bbf - \eta \jv)
$$
$$
= \int \Jbf \cdot (\vv \times \bv)dV - \int \eta |\jv|^2 dV -
\oint [p^1 + B \cdot \bv] \vv \cdot dS
$$
In deriving Eq. (6.1), we have used incompressibility. The first term can
cause transient growth while the Ohmic heating term is purely stabilizing.
The last term is the energy flux across the boundary and is zero by our
boundary conditions. 
For reduced MHD in a slab geometry with constant current,
$\Jbf (x) =\Jbf_0$, the energy transfer term,  $\int \Jbf \cdot (\vv \times \bv)dV $,
reduces to the Poisson bracket of the corresponding  flux functions,
$\Jbf_0  \int [\phi,\psi] dV $. In this case, the spatial integral  of the energy
transfer  vanishes (as shown by integrating by parts.) Thus no transient growth
occurs when $H(x)$ is linear. 

In the Alfv\'en wave heating problem, the antenna at the boundary sends a net
Poynting flux of energy into the plasma, and thereby 
forces the solution at the boundary.
In Ref.~6, Kappraff and Tataronis show that the Alfv\'en wave heating
problem has solutions which grow linearly in time until the Ohmic
dissipation saturates the growth.
Because the time integrated energy, which is transmitted by the antenna,
grows linearly in time, it is not
surprising that the kinetic energy grows initially as well.
Thus, both the initial value problem and the Alfv\'en heating problem
possess transiently growing solutions, but this growth is  surprising
for the initial value problem and is physically reasonable for  forced
problems such  as  Alfv\'en wave heating.




\section{\bf IX. SUMMARY}

The resistive magnetohydrodynamics operator is nonnormal and its eigenvalues
and eigenvectors are extremely sensitive to perturbation.
Using the WKBJ approximation, we have shown that
the entire stable half-annulus of complex frequencies with
$\rho{|\omega|^2}= |\kv \cdot \Bbf(x)|^2$ is
in the $\epsilon$-pseudospectrum and that the critical value of $\eps$
for these frequencies scales as  $\eps \sim \exp(-1/\eta^{1\over 2})$.
The frequency response Green's function is $O({1\over \eps})$
in this half-annulus, and thus, the finite time response is 
influenced by all of the frequencies in the half-annulus.

We believe that the resistive magnetohydrodynamic eigenfunctions form
a complete basis, but that this basis is so ill-conditioned,
$O(\exp(R_M^{1\over 2}))$, that
it is not useful  in describing the evolution of disturbances
on the ideal magnetohydrodynamic time-scale. From the scaling of the
condition number of the resistive eigenmode basis, we believe that the
eigenmode decomposition is  only relevant for times of order
$O(R_M^{1\over 2})$.

No transient growth occurs in a linear $\kv\cdot \Bbf (x)$ profile.
When the current density is not constant, our preliminary computations
indicate that weak transient amplification occurs.
When rational surfaces are present and
the initial perturbation has nonvanishing average of the normal magnetic
field perturbation on the rational surface (a tearing mode-like structure),
the ideal MHD pertubation grows linearly in time at the resonance surface. 
(See Appendix E.) However, the spatial extent of this perturbation 
may decay in time, and thus
the total energy of the perturbation need not  grow.

\

\noindent
{\bf Acknowledgement}

KSR acknowledges useful conversations with P.~Deift,  A.~Lifshitz,
L.~N.~Trefethen and H.~Weitzner, and especially E.~Hameiri
and S.~Reddy. We thank G.~Spies for allowing us to reproduce
Grad's proof of algebraic growth from Spies' lecture notes.
The manuscript has benefited from critical readings by
E.~Hameiri, S.~Reddy, and L.~N.~Trefethen.
The authors acknowledge L.~N.~Trefethen for providing preprints.
DNB acknowledges C.~A.~F.~Varandas for the support given during the realization
of this work.
KSR's work was performed under U.S.~Department of Energy, Grant No.
DE-FG02-86ER53223.

\section{\bf APPENDIX A: GENERALIZED EPSILON PSEUDOSPECTRA }

We now state the various equivalent definitions of the generalized
$\eps$-pseu\-do\-spec\-tra corresponding to the generalized eigenvalue problem,
$\Abf \ubf =\lambda \Mbf \ubf$.
We refer the reader to Ref.~16 for proofs of the equivalences.
We restrict our consideration to the finite dimensional case.
We denote the spectrum of the generalized eigenvalue problem,
${\bf A}{\bf e} = \lambda {\bf M}{\bf e}$, by
$\Lambda({\bf A}, {\bf M})$ and
the resolvent set by $\rho({\bf A},{\bf M})$.

\noindent
{Definition 2:}
 {\it Let $\Abf$ and $\Mbf$ be closed linear operators with domain $\DA$
and let $\Mbf$ be a positive definite self-adjoint operator such that
there exists a constant $c >0$ with $\Mbf \ge c\Ibf$.
Let $\epsilon \geq 0$ be given. A complex number $z$ is in the
$\epsilon$-pseudospectrum of $( {\bf A,M})$,
which we denote by $\Lambda_{\epsilon} ( {\bf A,M})$,
if any of the following equivalent conditions is satisfied:

(0) $\lambda$ is in the $\epsilon$-pseudospectrum of
$ \Fbf^{-*} {\bf AF}^{-1}$,
where $\Fbf^{*} \Fbf = {\bf M}$.

(i) the smallest generalized $({\bf M}^{-1}\ , \ \Mbf)$
singular value of ${\bf A}-\lambda{\bf M}$
is less than or equal to $\epsilon$,
i.e.~$\epsilon \ge min\{\mu({\bf A}-\lambda{\bf M},\Mbf^{-1},\Mbf) \}$.

(ii) there exists  $ {\bf u}\in \DA$ such that
${\bf u}^{*} {\bf {\bf Mu}} = 1$ and

\noindent
${\bf u}^{*} ({\bf A}-\lambda{\bf M})^{*}
{\bf M}^{-1} ({\bf A}-\lambda{\bf M}) {\bf u} \leq \epsilon^2$,

(iii) $\lambda$ is in the generalized spectrum of
${\bf A}+ \epsilon \Fbf^{*} {\bf E}\Fbf$:
$({\bf A} + \epsilon
\Fbf^{*} {\bf E}\Fbf) {\bf u} = \lambda{\bf M}{\bf u} $,
where $\Fbf^{*} \Fbf = {\bf M}$ and  ${\bf E}$ satisfies
$\parallel {\bf E} \parallel \leq 1$,

(iii') there exists an operator, ${\bf H}$, such that
$\lambda$ is in the generalized spectrum of ${\bf A}+ \epsilon {\bf H}:
({\bf A}+ \epsilon {\bf H}) {\bf u}
=\lambda{\bf Mu}$, where the matrix ${\bf H}$ satisfies
$$
\max_{{\bf u} \in C^n} {{\bf u}^{*} {\bf H}^{*}
{\bf M}^{-1} {\bf H}{\bf u} \over {\bf u}^{*} {\bf M}{\bf u}} \leq 1
\ \ . $$}

The equivalence of (i), (ii) and (iii) may be proved
by simply transforming each
of the properties from definition 1 to $\Fbf^{-*} {\bf AF}^{-1}$.

In our numerical computations, we use a different generalization of
$\epsilon$-pseu\-do\-spec\-tra. Our $\Mbf$-weighted $\epsilon$-pseudospectrum
has the advantage that definitions (i)-(iii) are simpler than in
Def.~2. However, the $\Mbf$-weighted $\epsilon$-pseudospectrum
is not related to the standard $\epsilon$-pseudospectrum
of Def.~1 through a change of variables. Thus the
$\Mbf$-weighted $\epsilon$-pseudospectrum is not based on the MHD energy norm,
and  Definition 3  has no analog of (0) in Def.~2.

\noindent
{Definition 3:}
{\rm ($\Mbf$-weighted $\epsilon$-pseudospectrum).} {\it
Let $\Abf$ and $\Mbf$ be closed linear operators with domain $\DA$
and let $\Mbf$ be a positive definite self-adjoint operator such that
there exists a constant $c >0$ with $\Mbf \ge c\Ibf$.
Let $\epsilon \geq 0$ be given and define $\epsbr \equiv \eps  ||\Mbf||$.
A complex number $\lambda$ is in the
$\Mbf$-weighted $\epsilon$-pseudospectrum of ${\bf A}$,
which we denote by $\Lambda_{\epsilon}( {\bf A} | \Mbf )$,
if one of the following equivalent conditions is satisfied:

(i) the smallest
singular value of ${\bf A}-\lambda{\bf M}$ is
less than or equal to $\epsbr$.

(ii) there exists ${\bf u}\in \DA$ such that
$||{\bf u}||^2 = 1$ and
$ ||({\bf A}-\lambda{\bf M}) {\bf u}||^2 \leq \epsbr^2$,

(iii) $\lambda$ is a generalized  eigenvalue of
${\bf A}+ \epsbr  {\bf E}$ w.r.t. {\bf M}:
$({\bf A} + \epsilon{\bf E}) {\bf u} = \lambda{\bf Mu} $,
 where the matrix ${\bf E}$ satisfies
$\parallel {\bf E} \parallel \leq 1$.
}

The normalization, $\epsbr \equiv \eps  ||\Mbf||$,
allows Def.~3 to reduce to Def.~1 when $\Mbf$ is a multiple of the
identity matrix.

\section{\bf APPENDIX B:  PHASE INTEGRAL FOR THE LINEAR PROFILE}

We evaluate the WKBJ phase function
for the  linear $H(x)={\kv \cdot \Bbf(x)}$ profile. The phase
integral becomes
$$\phi(x;\lambda)=\int _{{{{\lambda}}}
 - {{{{H(x_b)}}}}\over {{{{H(x_a)-H(x_b)}}}}}^{x}dx\,\sqrt{a x^2 + b x + c}=
$$
$$
\left[{{(2 {\sqrt{a}} b + 4 {a^{{3\over 2}}} x)  {\sqrt{\chi}}-{(b^2-4
a c)} \log ({b\over {2 {\sqrt{a}}}} + {\sqrt{a}} x +{\sqrt{\chi}}) \over
   {8 {a^{{3\over 2}}}}}}\right]_{{{{\lambda}}
 - {{{{H(x_b)}}}}\over {{{{H(x_a)-H(x_b)}}}}}}^{x}\ , \eqno(B1)$$
where
$$\chi\equiv a x^2 + b x + c={{{-{i} \over {\lambda}} \left({{-{{\lambda}^2} +
         {{\left( {{H(x_b)}} + {({x-x_a}) }
 \hskip1mm  {{{H(x_a)-H(x_b)}}\over {x_b-x_a}} \right)
}^2}}}\right)}}
\ .\eqno(B2)$$
The dispersion relation can be written in an
implicit form $F(\lambda)=\phi(r_b,\lambda)-\phi(r_a,\lambda)-n \pi=0$.
Newton iteration is used to solve the dispersion relation.
The derivative, $\pd(F,\lambda)$, is computed analytically using Eq.~(B1).
The analytic eigenvalues are given in Table 1.

Equation (3) gives a sufficient criterium for the validity of the WKBJ
expansion.
For the linear $H(x)$ profile, Eq.~(3)  reduces to
$${{({b \over {4 a}}  +
 {{x_{mn}+x_b-2 x_2}\over 2})
{\sqrt{\chi}}-{{(b^2-4a c)}\over
   {8 {a^{{3\over 2}}}}}
\log \left(\Pi\right)
 }}<0 \ ,\eqno(B3)$$
with
$$
\Pi={({b\over {2 {\sqrt{a}}}} + {\sqrt{a}} x_b +{\sqrt{\chi}})
({b\over {2 {\sqrt{a}}}} + {\sqrt{a}} x_{mn} +{\sqrt{\chi}})
\over ({b\over {2 {\sqrt{a}}}} + {\sqrt{a}} x_2 +{\sqrt{\chi}})^2}\ ,$$
and $x_{mn}$ satisfies $|H(x_{mn})| = |\lambda|$.

\section{\bf APPENDIX C:  FINITE-ELEMENT DISCRETIZATION}

In the Galerkin method,
a weak form of Eq.~(2) is constructed by multiplying the set
of equations with an arbitrary test function and integrating over the
domain of interest.
In this case, we use a finite-element basis with
the actual functions as test functions. We rewrite Eq.~(2)
in the reduced MHD form:
$$
\lambda {\nabla_\bot^2 U_1}={\bf B}_0 \cdot \nabla (\nabla_\bot^2
A_1) + {(\nabla A_1 \times \hat z)} \cdot \nabla j_z,
$$
$$\lambda A_1 = {\bf B}_0 \cdot \nabla (U_1)+ \eta \nabla_\bot^2 A_1,\eqno(C1)$$
where $A,U$ are the stream
functions defined as,
${\bf B_1} =\nabla A_1 \times \hat z + {\bf B}_z \hat z,$
${\bf v}=\nabla U_1 \times \hat z,$ and $\nabla_\bot =\nabla - \hat z \pd(,z),$
the resulting generalized eigenvalue
problem, $\Abf {\bf u}=\lambda \Mbf {\bf u}$, has
matrix elements:      

$$\tens A(A_1,A_1)=-\eta \int \nabla_\bot A_1^* \cdot \nabla_\bot A_1 dV +\eta \int A_1^*
\nabla_\bot A_1\cdot \vec {\bf n} dS, $$

$$\tens A(A_1,U_1)=\int A_1^* ({\bf B}_0 \cdot \nabla) U_1 dV. $$

$$\tens A(U_1,A_1)=-\int U_1^* ({\bf B}_1 \cdot \nabla) j_z dV +\int U_1^* ({\bf B}_0
\cdot \nabla) \nabla_\bot^2 A_1 dV $$
$$=-\int U_1^* ({\bf B}_1 \cdot \nabla) j_z dV-\int \nabla_\bot U_1^*
\cdot ({\bf B}_0 \cdot \nabla) \nabla_\bot A_1 dV $$

$$-\int U_1^* \nabla_\bot ({\bf B}_0 \cdot \nabla) \nabla_\bot A_1
dV+\int U_1^* ({\bf B}_0 \cdot \nabla) \nabla_\bot A_1 \cdot n ds,
$$

$$\tens M(A_1,A_1)=\int A_1^* A_1 dV ,$$

$$\tens M(U_1,U_1)=-\int \nabla_\bot U_1^* \cdot \nabla_\bot U_1 dV + \int U_1^*
\nabla_\bot U \cdot \vec {\bf n} dS .\eqno(C2)$$

$\Mbf$ is a Hermitian, positive-definite matrix.
Due to the local support of the finite-elements, the integrand is nonzero only
for neighboring points. A more detailed description of the numerical
discretization is contained in Ref.~23.

For generalized eigenvalue problems, the QZ algorithm is usually recommended.
However, we found that this algorithm is not stable for
this kind of matrices. The error
propagation is too large and the results are contaminated.
The best results were
obtained by inverting the $\Mbf$ matrix and solving the eigenvalue problem
$\Mbf^{-1} \Abf \uv= \lambda \uv$ applying the QR algorithm.
\section{\bf APPENDIX D: ANALYTIC EPSILON-PSEUDOSPECTRUM}

To calculate the $\eps$-pseudospectrum analytically, we neglect the
$H''(x)$ term in Eq.~(2a). By dropping this term, the WKBJ solutions
decouple from the ideal MHD solutions and the RMHD operator,
$(\Lbf - \lambda) \psi$ reduces to
a second order equation:
$$\Lbf_{\lambda} \psi =
\eta \nabla^2 \psi - i \lambda \left[1 - {H(x)^2
\over\lambda^2} \right] \psi \ ,\eqno(D1)$$
where $\lambda$ is now a nonlinear eigenvalue parameter.
In general, this simplification is not valid for $\lambda$ values
which have two anti-Stokes line crossing the interval, $[x_a,x_b]$.
(See Figure 1.) In this $\lambda$-region, the ideal solutions couple to
the WKBJ solutions. For $\lambda$ which have
at most one anti-Stokes line crossing $[x_a,x_b]$, Ref.~9 shows that the
formal solutions do not couple. In this case, our analysis of the
$\eps$-pseudospectrum will be valid if the the $\eps$-pseudomode oscillates
rapidly on the scale length of the WKBJ solutions.

To determine the $\eps$-pseudospectrum, we construct the Green's function
for

\noindent
Eq.~(D1) using the WKBJ solutions:
$$\Psi_{\pm}(x) = {(H(x)^2-\lambda^2)}^{-{1 \over 4}}e^{ \pm i \phi (x)}
...\eqno(D2)$$
The Green's function, $G(r,x)$, satisfies
$$\eta \nabla_r^2 G(r,x) - i \lambda \left[1 - {H(r)^2
\over \lambda^2}\right] G(r,x)=\delta(r-x),\eqno(D3)$$
with the boundary conditions: $G(x_a,x)= 0$ and $G(x_b,x)= 0$.
We define the function
$\chi(y,z)\equiv{\Psi_+(y) \Psi_-(z) - \Psi_+(z) \Psi_-(y)}$, and the functions,
$\Psi_0(r,x)$ and $\Psi_1(r,x)$:
$$\Psi_0(r,x)= { \chi(x_a,r)\chi(x_b,x) \over  \chi(x_b,x_a) }
\ ,\eqno(D4)$$
$$
\Psi_1(r,x)= { \chi(x_b,r)\chi(x_a,x) \over  \chi(x_b,x_a) }
\ .\eqno(D5)$$
Note $\Psi_0(x_a,x)=0$, $\Psi_1(x_b,x)=0$ and $\Psi_0(x,x)=\Psi_1(x,x)$
at $r= x$. At
$r= x$, the first derivatives satisfy the
jump condition:
$\partial_r\Psi_0(x,x)=\partial_r \Psi_1(x,x)+1$.
Furthermore,
$$ \Psi_1(r,x) - \Psi_0(r,x) =
{\Psi_+(r)\Psi_-(x)}-{\Psi_-(r) \Psi_+(x)} = \chi(r,x) .$$
Thus, the Green's function can be rewritten as, $$G(r,x)=\Psi_0(r,x) +
\chi(r,x) \ \Theta(r-x),\eqno(D6)$$
where $\Theta$ is the Heaviside function.
The resolvent, $||{(\lambda \tens I-\Lbf)}^{-1}||$, is represented as
$${(\lambda \tens I-\Lbf)}^{-1} f = \int_{x_a}^{x_b}G(r,x)f(x)dx .\eqno(D7)$$

To calculate the $\eps$-pseudospectrum, we determine the norm of
the resolvent by
maximizing $||{(\lambda \tens I-\Lbf)}^{-1}f|| / ||f||$.
When the WKBJ expansion is valid,
$\chi(y,z) \sim O(\exp( i \int_y^z \phi'(x,\lambda)dx))$.
Since ${ \chi(x_a,r)\chi(x_b,x) >>  \chi(r,x) \chi(x_b,x_a) }$,
we neglect

\noindent $\chi(r,x)\Theta(r-x)$ in Eq.~(D6).
Thus ,
$$||{(\lambda \tens I-\Lbf)}^{-1}||\simeq{{Sup} _{f(x)}}
\left|\left|\int_{x_a}^{x_b}dx\, {\chi(x_a,r)\chi(x_b,x)f(x)
\over { \chi(x_b,x_a) ||f||}} \right|\right| \ .\eqno(D8)$$
$\Psi_0(r,x)$ and $\Psi_1(r,x)$ are largest at $r= x_{mn}= x$.
The supremum occurs for $f(x) = \chi(x_b,x)$.
The resulting expression for  the $\epsilon$-
pseudospectrum  is
$${\epsilon}_{bd}(\lambda) \approx {{\chi(x_a,x_b)}
\over {\chi(x_b,x_{max})
\chi(x_a,x_{max})}} \ .\eqno(D9)$$
This analysis is only valid when the formal WKBJ solutions are valid and
Im$[\phi'(x,\lambda)]$ has it minimum in the interval.
(Property 1 shows that $x_{mn}$ satisfies
$\rho | \lambda |^2 = H(x_{mn})^2$.)
Our analysis of the $\epsilon$-pseudospectrum is based on similar analysis
for the convection diffusion problem given in Ref.~17.

\section{\bf APPENDIX E: ALGEBRAIC GROWTH IN IDEAL MHD}

We now present a result of H.~Grad's which shows that the linearized
circulation grows algebraically in time in ideal MHD
for certain perturbations. Equations (E1)-(E2) are
are from Ref.~2. We consider a closed flux  line and
define the first order circulation as
$$
c(t) = \oint {\bf d} \ell \cdot {\bf u}_1 \ , 
$$
where the contour integral is along the field line.
We now evaluate ${dc \over dt}$ and ${d^2c \over dt^2}$.
Since the equilibrium is static, the path of integration does not move to
lowest order, and the time derivative of $c$ is obtained by just
differentiating the integrand. Using the linearized equations of motion and
the equilibrium equations and  assuming that $\nabla \rho_0 \times
\nabla p_0 = 0$, we compute
$${dc \over dt} = \oint {\bf d} \ell \cdot {\partial {\bf u}_1 \over
\partial t}\
 =\  {1 \over \rho_0 } \oint {\bf d} \ell \cdot (- \nabla p_1 +
{\bf j} \times {\bf B} + {\bf J} \times {\bf b} ) $$
$$  =\  {1 \over \rho_0} \oint {\bf d} \ell \cdot ({\bf J} \times
{\bf b} )
  =\  {1 \over \rho_0 } \oint  d
\ell {1 \over  |B|} {\bf B} \cdot ({\bf J} \times {\bf b} ) $$
$$  =\  - {1 \over \rho_0 } \oint d \ell {1 \over |B|} {\bf b} \cdot
({\bf J} \times {\bf B} )
  =\  - {1 \over \rho_0} \oint d \ell {1 \over |B|} ({\bf b} \cdot
\nabla p_0 )
\eqno (E1)$$

and further
$$      
{d^2 c \over dt^2} = - {1 \over \rho_0}
\oint  d \ell  {1\over B} \left( {\partial {\bf b} \over
\partial t} \cdot \nabla p_0 \right)
 =\  - {1 \over \rho_0}
 \oint  d \ell {1 \over B} \nabla p_0 \cdot {\rm curl}
({\bf u} \times {\bf B} ) $$
$$  =\  - {1 \over \rho_0} \oint  d \ell {1 \over |B| } \ {\rm div}
(({\bf u} \times
{\bf B} )  \times \nabla p_0 )
  =\  - {1 \over \rho_0 } \oint  d
\ell {1 \over  |B|} {\rm div} (({\bf u} \cdot \nabla p_0 )
{\bf B} ) $$
$$  =\  - {1 \over \rho_0 } \oint
{\bf d}\ell {1 \over |B|} {\bf B} \cdot \nabla
({\bf u} \cdot \nabla p_0 )
  =\  - {1 \over \rho_0} \oint
{\bf d}\ell  \cdot \nabla ( {\bf u}_1 \cdot \nabla p_0 ) = 0
\ .\eqno(E2)$$
Hence $dc/dt$ is constant. When ${\bf b}$ has the form:
${\bf b} = {\rm curl} [ {\bf \xi} \times {\bf B}]$ with a
single-valued vector field ${\bf \xi}$, then $dc/dt = 0$.
(This is shown by using vector identities similar to those in Eq.~(E2).)
Since $c(t)$ is growing linearly, the maximum of $\uv({\bf x},t)$ is growing at least
linearly.

In slab geometry with a single helicity perturbation:
$\bv({\bf x},t)= e^{i (k z + m y )} \bv({x})$, the transient
growth criterion of  Eq.~(E1) reduces to 
$b_x (x_{res} ) \neq 0$, i.e.
the perturbed normal flux at the resonance surface does not average to zero.

We strengthen Grad's result by noting that the transiently growing ideal
MHD solution is approximately a solution of the RMHD equations for small
enough resistivity. Thus the RMHD equations will have transient growth
in the supremum norm for tearing mode perturbations.
Pointwise growth of the perturbation does not imply growth in the energy
norm because the spatial extent of the  perturbation can decrease. For the
case of a linear profile, $H(x) =x$, Sec.~VIII shows that this profile is stable
in the energy norm while growing in the supremum norm. 

Landahl$^{24}$ has shown that the
inviscid Orr-Sommerfeld equation has solutions which grow linearly in time.
Landahl's unstable modes are global modes while the circulation instability
of Eqs.~(E1)-(E2) is localized on a field line.

\newpage
\section{\bf References}


1. I.~Bernstein, E.~A.~Freeman, M.~D.~Kruskal, R.~M.~Kulsrud,
Proc.~Royal~Soc. Ser.~A {\bf 244}, 17 (1958).

2. G.~O.~Spies, 
National Technical Information Service document No. COO-3077-137,
``Elements of magnetohydrodynamic stability theory,'' New York
University Report MF-86, November 1976. Copies may be ordered from
the National Technical Information Service, Springfield, Virginia
22161. 

``Elements of magnetohydrodynamic stability theory,''
{Courant Institute of Mathematical Sciences Report MF-86}
New York Univiersity, (1976).

3. W.~Grossmann and J.A.Tataronis, Z.~Phys.  {\bf 261}, 217 (1973).


4. E.~Hameiri, Comm.~Pure \& Applied Math, {\bf 38}, 43 (1985).

5. P.~Laurence,  J.~Math.~Phys. {\bf 27}, 1916 (1986).

6. J.~M.~Kappraff and J.A.~Tataronis, J.~Plasma Phys. {\bf 18}, 209 (1977).

7. C.~M.~Ryu and R.~C.~Grimm, J.~Plasma Phys. {\bf 32}, 207 (1984).

8. Y.~Pao and W.~Kerner, Phys.~Fluids, {\bf 28}, 287 (1985).

9. K.~S.~Riedel, Phys.~Fluids.~{\bf 29}, 1093 (1986).

10. W.~Kerner, K.~Lerbinger, K.S.~Riedel,
Phys.~Fluids, {\bf 29}, 2975  (1986).

11. D.~Lortz and G.~O.~Spies, Physics Letters {\bf 101A}, 335 (1984).

12. L.~N.~Trefethen, ``Pseudospectra of matrices,'' in
Numerical Analysis 1991,
edited by D.~F.~Griffiths and
G.~A.~Watson, (Longman Scientific and Technical Press, Harlow, UK 1992).

13. S.~C.~Reddy, P.~J.~Schmid, and D.~S.~Henningson, SIAM J.~Appl.~Math.
{\bf 53}, 15 (1993).

14. S.~C.~Reddy and D.~S.~Henningson, J.~Fluid~Mech.
{\bf 252}, 209  (1993).

15. L.~N.~Trefethen, A.~E.~Trefethen, S.~C.~Reddy, and T.~A.~Driscoll,
Science {\bf 261}, 578 (1993).

16. K.~S.~Riedel, ``Generalized epsilon pesudospectra,'' to be published in
SIAM J.~Numerical Anal.~{\bf 31}, (August, 1994).

17. S.~C.~Reddy, and L.~N.~Trefethen,  ``Pseudospectra of the convective
diffusion equation,''
to be published in SIAM J.~Appl.~Math.
(1994).

18. R.C.~Diprima and G.~J.~Habetler, Arch.~Rat.~Mech.~Anal.
{\bf 34}, 218 (1969).

19. K.~Butler and B.~Farrell, Phys.~Fluids A~{\bf 4}, 8 (1992).

20. A.~W.~Naylor and G.R.~Sell, Linear Operators in the Engineering Sciences
(Springer Verlag, New York 1982) p.~412.

21. F.~Riesz and B.~Sz.-Nagy, Functional Analysis
(F.~Ungar Publishing Company, New York 1955) p.~364.

22. S.~Poedts and W.~Kerner, J.~Plasma Physics {\bf 47}, 139 (1992) .

23. W.~Kerner, K.~Lerbinger, R.~Gruber, T.~Tsunematsu,
Comp.~Phys.~Comm.

\noindent {\bf 36}, 225 (1985).

24. M.~T.~Landahl, J.~Fluid~Mech.~{\bf 98}, 243 (1980).

\includepdf[pages=-,pagecommand={}]{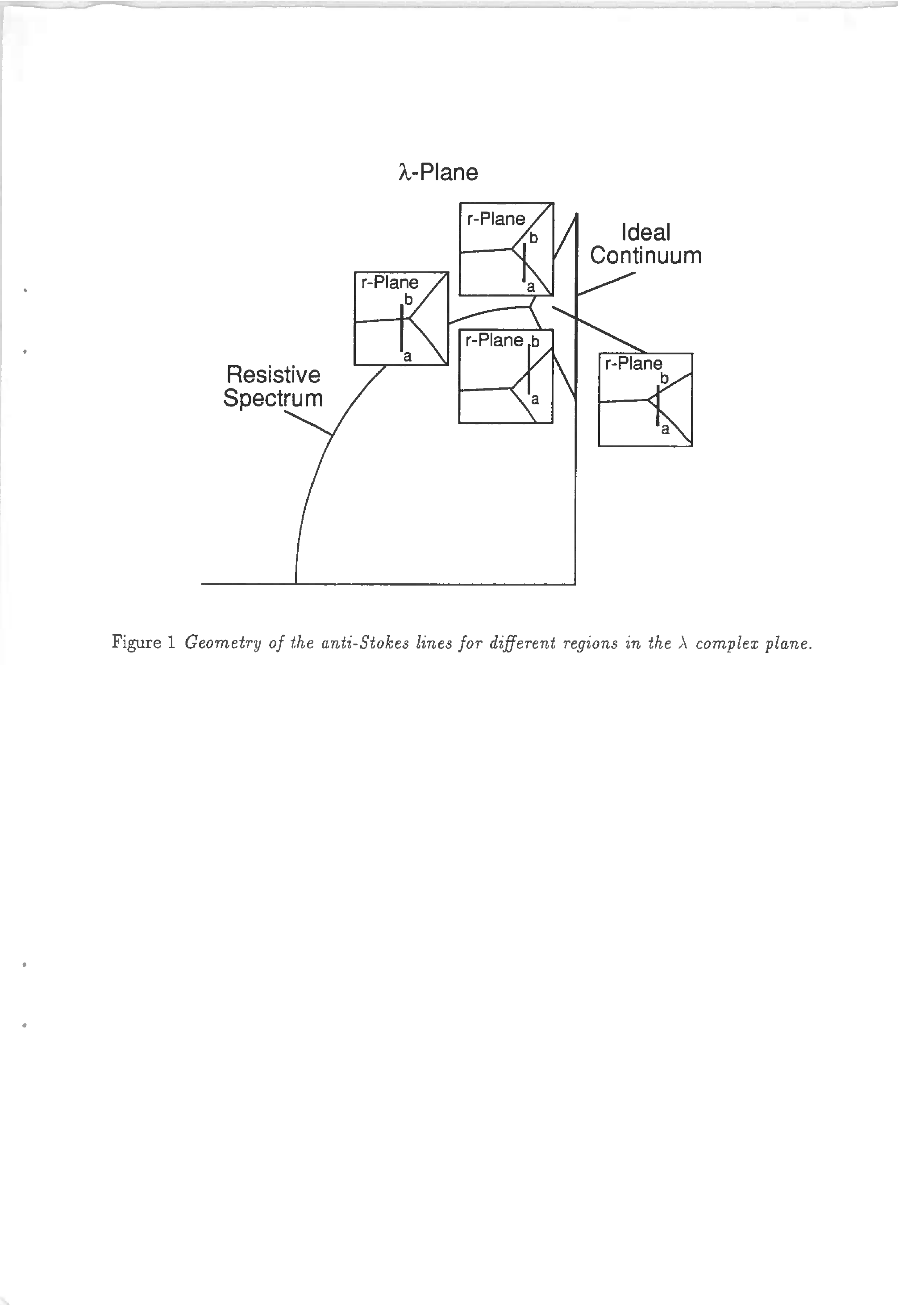}
\end{document}